\documentclass[aps,twocolumn,showpacs]{revtex4}
\begin{document}
\title{Nonrelativistic Quantum Mechanics with Spin \\
in the Framework of a Classical Subquantum Kinetics}
\author{G. Kaniadakis}
\affiliation{Dipartimento di Fisica and Istituto Nazionale di
 Fisica della Materia, \\ Politecnico di Torino,
Corso Duca degli Abruzzi 24, 10129 Torino, Italy \\
E-mail address: kaniadakis@polito.it}
\date{\today}

\begin{abstract}
Recently it has been shown that the spinnless one particle
quantum mechanics can be obtained in the framework of entirely
classical subquantum kinetics. In the present paper we argue that,
within the same scheme and without any extra assumption, it is
possible to obtain both the non relativistic quantum mechanics
with spin, in the presence of an arbitrary external
electromagnetic field, as well as  the nonlinear quantum
mechanics.
\end{abstract}
\maketitle

\section{Introduction}

Certainly one of the most intricate questions in quantum
mechanics concerns the origin of quantum potential
\cite{MA,BO,RE,GK}. Recently in ref. \cite{GK} the quantum
potential has been obtained in the framework of classical many
body physics without invoking any epistemological principle. This
unexpected result gives rise to the hope that quantum physics can
be obtained in a self consistent scheme of an entirely classical
many body physics. Indeed, in the same reference within this
classical framework, the main features of quantum mechanics (i.e.
the probabilistic nature of the quantum description, the
Schr\"odinger equation, the quantum operators, the Heisenberg
uncertainty principle) have been deduced in the case of a spinless
particle. It is surprising that this theory contains only one
free parameter which is identifiable with the fundamental
constant $\hbar$.

Within this classical scheme the quantum particle turns out to
have an internal structure and a spatial dispersion and appears
to be composed by N identical point like and interacting
subquantum objects, the monads, obeying to the laws of classical
physics. The statistical ensemble of these $N$ monads is governed
in the phase space by a kinetics, which takes into account that,
during the point-like collisions, the monad number, momentum and
energy are conserved. It is remarkable that simply the projection
of the above phase space kinetics, into the physical space,
produces a hydrodynamics which leads naturally to the one
particle spinless quantum mechanics, when the external force field
is conservative.

Clearly this latter condition is instead restrictive. A first
question which naturally arises  is if it is possible to obtain
quantum mechanics for a particle with spin in the presence of an
arbitrary time dependent external electromagnetic field
\cite{TA,SO,SI,GH,WBZ}, within the above $N$-body classical
kinetics. A second question is if it is possible to obtain also
the nonlinear quantum mechanics \cite{KS}, within the same scheme.
The answers to these questions, as we will see in the following,
are affirmative.

The main purpose of the present work is to show that it is
possible to obtain both the non relativistic quantum mechanics
with spin as well the nonlinear quantum mechanics, following the
general lines of the theory developed in ref. \cite{GK}, without
making any additional assumption.

\section{Classical Hydrodynamics}

Let us consider the phase space kinetics of a system of $N$
monads, each of mass $\mu$ and charge $q$. We suppose that the
system of total mass $m=N\mu$ and charge $e=Nq$, is immersed in an
arbitrary external electromagnetic field derived from the time
dependent potential $A ^{^{{\scriptscriptstyle(}{\scriptstyle
ex}{\scriptscriptstyle)}}}\!\!=\!\!\Big(A_0
^{^{{\scriptscriptstyle(}{\scriptstyle ex}{\scriptscriptstyle)}}}
\!\!,\mbox{\boldmath ${A}$}
^{^{{\scriptscriptstyle(}{\scriptstyle ex}{\scriptscriptstyle)}}}
\Big)$ by means of
\begin{equation}
\mbox{\boldmath ${E}$} ^{^{{\scriptscriptstyle(}{\scriptstyle
ex}{\scriptscriptstyle)}}}\!\!=\!- \mbox{\boldmath $\nabla$} A_0
^{^{{\scriptscriptstyle(}{\scriptstyle
ex}{\scriptscriptstyle)}}}\!-\!\frac{1}{c}\frac{\partial
\mbox{\boldmath ${A}$} ^{^{{\scriptscriptstyle(}{\scriptstyle
ex}{\scriptscriptstyle)}}}}{\partial t} \ \ ; \ \ \mbox{\boldmath
${B}$} ^{^{{\scriptscriptstyle(}{\scriptstyle
ex}{\scriptscriptstyle)}}}\! \!=\! \mbox{\boldmath $\nabla$}
\times \mbox{\boldmath ${A}$}
^{^{{\scriptscriptstyle(}{\scriptstyle ex}{\scriptscriptstyle)}}}
\ . \label{II1}
\end{equation}
The distribution function, obeying the normalization condition
$\int f d^{3}v d^{3} x=N $ of such a system, is governed by the
kinetic equation \cite{LI,GL}
\begin{equation}
\frac{\partial {f}}{\partial t}
 + \mbox{\boldmath $v$}
\cdot \mbox{\boldmath $\nabla$}f + \frac{q}{\mu}
\left(\mbox{\boldmath ${E}$}
^{^{{\scriptscriptstyle(}{\scriptstyle
ex}{\scriptscriptstyle)}}}\!\!+\!\frac{1}{c}\,\mbox{\boldmath
${v}$} \times\mbox{\boldmath ${B}$}
^{^{{\scriptscriptstyle(}{\scriptstyle
ex}{\scriptscriptstyle)}}}\right)\cdot \mbox{\boldmath
$\nabla$}_{_ {\!\! \scriptstyle v}}\, f =C({ f})\ \ . \label{II2}
\end{equation}

The assumption that during the point-like collisions the monad
number, momentum and energy are conserved,  implies that the three
functions $g_1(\mbox{\boldmath $v$})=1$, $g_2(\mbox{\boldmath
$v$})=\mbox{\boldmath $v$}$ and $g_3(\mbox{\boldmath
$v$})=\mbox{\boldmath $v$}^2$ are the collision invariants of the
system and thus the collision integral $C(f)$ satisfies the
conditions: $\int g_j(\mbox{\boldmath $v$})C(f) d^{3} v=0$ with
$j=1,2,3$.

In the following we study the projection of the system dynamics
in the physical space, where the distribution function is
$\rho(t, \mbox{\boldmath $x$})= \int f d^{3} v $ and the mean
value of a given property $G=G(t, \mbox{\boldmath $x$} ,
\mbox{\boldmath $v$})$ of the system, in the point $
\mbox{\boldmath $x$}$, is defined as $ <\!G\!> \!\!_{v}
=\rho^{-1}\int G \,f\, d^{3}v$. Then, we can introduce the density
of current $\mbox{\boldmath $u$}= <\mbox{\boldmath $v$}>\!\!_{v}$
and the symmetric tensor $ \sigma_{ij} = \sigma_{ji} $ density of
stress
\begin{eqnarray}
\sigma_{jk}\,=\mu\,\big(\!<v_j\,v_k> \!\!_{v}-<v_j> \!\!_{v}
<v_k> \!\!_{v}\big)
 \ \ .\label{II4}
\end{eqnarray}

Multiplying Eq.(\ref{II2}) by the two first collision invariants
$g_1(\mbox{\boldmath $v$})$ and $g_2(\mbox{\boldmath $v$})$ and
after integration with respect to $\mbox{\boldmath $v$}$, the
continuity equation
\begin{equation}
\frac{\partial \rho}{\partial t} +\mbox{\boldmath $\nabla$}\!
\cdot (\rho\mbox{\boldmath $u$})=0\ \ , \label{II5}
\end{equation}
and the momentum balance equation
\begin{equation}
\!\!\!\frac{\partial}{\partial t}(\mu \rho u_j)\! +\!
\frac{\partial }{\partial x_k}(\mu \rho u_ju_k+\rho\sigma_{jk})\!
-\! \rho {F}_j ^{^{{\scriptscriptstyle(}{\scriptstyle
ex}{\scriptscriptstyle)}}}\!\!\!=\!0 \ , \label{II6}
\end{equation}
with
\begin{equation}
\mbox{\boldmath ${F}$} ^{^{{\scriptscriptstyle(}{\scriptstyle
ex}{\scriptscriptstyle)}}}=q \left(\mbox{\boldmath ${E}$}
^{^{{\scriptscriptstyle(}{\scriptstyle
ex}{\scriptscriptstyle)}}}\!\!+\!\frac{1}{c}\,\mbox{\boldmath
${u}$} \!\times\!\mbox{\boldmath ${B}$}
^{^{{\scriptscriptstyle(}{\scriptstyle
ex}{\scriptscriptstyle)}}}\right) \ , \label{II7}
\end{equation}
can be obtained respectively \cite{LI,GL}.

It is important to emphasize that in the Lorentz force acting on a
single monad, as we can see from Eq. (\ref{II2}), appears the
monad velocity $\mbox{\boldmath $v$}$, while in the expression of
the Lorentz $\mbox{\boldmath ${F}$}
^{^{{\scriptscriptstyle(}{\scriptstyle
ex}{\scriptscriptstyle)}}}$ given by Eq. (\ref{II7}), appears
$\mbox{\boldmath $u$}$ which describes a collective property of
the system. Eq.s (\ref{II5}) and (\ref{II6}) are the hydrodynamic
equations for the system, which conserves its monad number $N=\int
\rho\, d^{3} x$ and behaves as a fluid in the physical space.

After taking into account Eq. (\ref{II5}), we rewrite Eq.
(\ref{II6}) in the following Newton-like form
\begin{eqnarray}
\mu\frac{D \mbox{\boldmath $u$} }{D t}= \mbox{\boldmath ${F}$}
^{^{{\scriptscriptstyle(}{\scriptstyle
\sigma}{\scriptscriptstyle)}}}+ \mbox{\boldmath ${F}$}
^{^{{\scriptscriptstyle(}{\scriptstyle ex}{\scriptscriptstyle)}}}
\ \ , \label{II8}
\end{eqnarray}
where $D/D t=\partial/\partial t + \mbox{\boldmath $u$}\cdot\,
\mbox{\boldmath $\!\!\nabla$}$ is the total time (or Lagrangian or
substantial) derivative. The hydrodynamic force $\mbox{\boldmath
$F$}^{(\sigma)}\!$, generated, through the stress tensor, from the
monad interactions, takes the form
\begin{equation}
{F}_j ^{^{{\scriptscriptstyle(}{\scriptstyle
\sigma}{\scriptscriptstyle)}}}=-\frac{1}{\rho}\,\frac{\partial
}{\partial x_k} \,\rho\,\sigma_{jk} \ \ . \label{II10}
\end{equation}

We now focus our attention to an interesting property of the force
$-\mu D \mbox{\boldmath $u$}/Dt$. It is trivial to verify that
this force has a Lorentz-like structure
\begin{equation}
- \mu\frac{D \mbox{\boldmath $u$} }{D t}=q \left(\mbox{\boldmath
${E}$} ^{^{{\scriptscriptstyle(}{\scriptstyle
u}{\scriptscriptstyle)}}}\!\!+\!\frac{1}{c}\,\mbox{\boldmath
${u}$} \!\times\!\mbox{\boldmath ${B}$}
^{^{{\scriptscriptstyle(}{\scriptstyle
u}{\scriptscriptstyle)}}}\right) \ , \label{II11}
\end{equation}
where the fields $\mbox{\boldmath ${E}$}
^{^{{\scriptscriptstyle(}{\scriptstyle u}{\scriptscriptstyle)}}}$
and $\mbox{\boldmath ${B}$} ^{^{{\scriptscriptstyle(}{\scriptstyle
u}{\scriptscriptstyle)}}}$ can be derived from the potential $A
^{^{{\scriptscriptstyle(} {\scriptstyle u}{\scriptscriptstyle)}}}$
\begin{equation}
A ^{^{{\scriptscriptstyle(}{\scriptstyle
u}{\scriptscriptstyle)}}}=\left(A_0
^{^{{\scriptscriptstyle(}{\scriptstyle u}{\scriptscriptstyle)}}}
,\mbox{\boldmath ${A}$} ^{^{{\scriptscriptstyle(}{\scriptstyle
u}{\scriptscriptstyle)}}}
\right)=\frac{\mu}{q}\left(\frac{1}{2}\mbox{\boldmath$u$}^2,
c\mbox{\boldmath$u$}\right) \ \ , \label{II12}
\end{equation}
according to
\begin{equation}
\mbox{\boldmath ${E}$} ^{^{{\scriptscriptstyle(}{\scriptstyle
u}{\scriptscriptstyle)}}}\!=\!-\mbox{\boldmath $\nabla$} A_0
^{^{{\scriptscriptstyle(}{\scriptstyle
u}{\scriptscriptstyle)}}}\!-\frac{1}{c}\frac{\partial
\mbox{\boldmath ${A}$} ^{^{{\scriptscriptstyle(}{\scriptstyle
u}{\scriptscriptstyle)}}}}{\partial t} \ \ \ ; \ \ \
\mbox{\boldmath ${B}$} ^{^{{\scriptscriptstyle(}{\scriptstyle
u}{\scriptscriptstyle)}}} \!=\!\mbox{\boldmath $\nabla$} \times
\mbox{\boldmath ${A}$} ^{^{{\scriptscriptstyle(}{\scriptstyle
u}{\scriptscriptstyle)}}} \ \ . \label{II13}
\end{equation}
The Lorentz-like structure of $-\mu D \mbox{\boldmath $u$}/Dt$ is
exclusively enforced by the projection mechanism of particle
motion from the phase space into the physical space (kinetics
$\rightarrow$ hydrodynamics). An immediate consequence of the
Lorentz-like structure of $-\mu D \mbox{\boldmath $u$}/Dt$ is that
Eq. (\ref{II8}) imposes a Lorentz-like structure also for
$\mbox{\boldmath ${F}$} ^{^{{\scriptscriptstyle(}{\scriptstyle
\sigma}{\scriptscriptstyle)}}}$:
\begin{equation}
\mbox{\boldmath ${F}$} ^{^{{\scriptscriptstyle(}{\scriptstyle
\sigma}{\scriptscriptstyle)}}}=q \left(\mbox{\boldmath ${E}$}
^{^{{\scriptscriptstyle(}{\scriptstyle
\sigma}{\scriptscriptstyle)}}}\!\!+\!\frac{1}{c}\,\mbox{\boldmath
${u}$} \!\times\!\mbox{\boldmath ${B}$}
^{^{{\scriptscriptstyle(}{\scriptstyle
\sigma}{\scriptscriptstyle)}}}\right) \ , \label{II14}
\end{equation}
where the fields $\mbox{\boldmath ${E}$}
^{^{{\scriptscriptstyle(}{\scriptstyle
\sigma}{\scriptscriptstyle)}}}$ and $\mbox{\boldmath ${B}$}
^{^{{\scriptscriptstyle(}{\scriptstyle
\sigma}{\scriptscriptstyle)}}}$ can be derived from the potential
$A ^{^{{\scriptscriptstyle(} {\scriptstyle
\sigma}{\scriptscriptstyle)}}}=(A_0
^{^{{\scriptscriptstyle(}{\scriptstyle
\sigma}{\scriptscriptstyle)}}} ,\mbox{\boldmath ${A}$}
^{^{{\scriptscriptstyle(}{\scriptstyle
\sigma}{\scriptscriptstyle)}}})$ by means of
\begin{equation}
\mbox{\boldmath ${E}$} ^{^{{\scriptscriptstyle(}{\scriptstyle
\sigma}{\scriptscriptstyle)}}}\!\!=-\mbox{\boldmath $\nabla$} A_0
^{^{{\scriptscriptstyle(}{\scriptstyle
\sigma}{\scriptscriptstyle)}}}\!-\frac{1}{c}\frac{\partial
\mbox{\boldmath ${A}$} ^{^{{\scriptscriptstyle(}{\scriptstyle
\sigma}{\scriptscriptstyle)}}}}{\partial t} \ \ ; \ \
\mbox{\boldmath ${B}$} ^{^{{\scriptscriptstyle(}{\scriptstyle
\sigma}{\scriptscriptstyle)}}}\!\! = \mbox{\boldmath $\nabla$}
\times \mbox{\boldmath ${A}$}
^{^{{\scriptscriptstyle(}{\scriptstyle
\sigma}{\scriptscriptstyle)}}} \ \ . \label{II15}
\end{equation}

We observe now that, after introducing the two fields
$\mbox{\boldmath ${E}$}=\mbox{\boldmath ${E}$}
^{^{{\scriptscriptstyle(}{\scriptstyle
u}{\scriptscriptstyle)}}}\!\!+\!\mbox{\boldmath ${E}$}
^{^{{\scriptscriptstyle(}{\scriptstyle
\sigma}{\scriptscriptstyle)}}}\!\!+\!\mbox{\boldmath ${E}$}
^{^{{\scriptscriptstyle(}{\scriptstyle ex}{\scriptscriptstyle)}}}$
and $\mbox{\boldmath ${B}$}=\mbox{\boldmath ${B}$}
^{^{{\scriptscriptstyle(}{\scriptstyle
u}{\scriptscriptstyle)}}}\!\!+\!\mbox{\boldmath ${B}$}
^{^{{\scriptscriptstyle(}{\scriptstyle
\sigma}{\scriptscriptstyle)}}}\!\!+\!\mbox{\boldmath ${B}$}
^{^{{\scriptscriptstyle(}{\scriptstyle
ex}{\scriptscriptstyle)}}}$, which can be derived from the
potential $A=(A_0,\mbox{\boldmath
${A}$})=A^{^{{\scriptscriptstyle(}{\scriptstyle
u}{\scriptscriptstyle)}}}\!\!+\!
A^{^{{\scriptscriptstyle(}{\scriptstyle
\sigma}{\scriptscriptstyle)}}}\!\!+\!
A^{^{{\scriptscriptstyle(}{\scriptstyle
ex}{\scriptscriptstyle)}}}$, Eq. (\ref{II8}) can be written as
\begin{equation}
\mbox{\boldmath ${E}$} + \,\frac{1}{c}\,\mbox{\boldmath ${u}$}
\!\times \mbox{\boldmath ${\!B}$}=0 \ . \label{II16}
\end{equation}
Of course, Eq. (\ref{II16}) is satisfied if simultaneously
$\mbox{\boldmath ${E}$}=0 $ and $\mbox{\boldmath ${B}$}=0 $. The
condition $\mbox{\boldmath ${B}$}=0$ implies $\mbox{\boldmath
$\nabla$} \times \mbox{\boldmath ${A}$}=0$ so that we can write
$q \mbox{\boldmath ${A}$}= c\mbox{\boldmath $\nabla$}{\cal S}$.
From this last relation and the definition of $\mbox{\boldmath
${A}$}$ we obtain
\begin{equation}
\mbox{\boldmath ${u}$}=\frac{1}{\mu}\left( \mbox{\boldmath
$\nabla$} {\cal S }-\frac{q}{c}\mbox{\boldmath ${A}$}
^{^{{\scriptscriptstyle(}{\scriptstyle
ex}{\scriptscriptstyle)}}}\!\!\!-\frac{q}{c}\mbox{\boldmath ${A}$}
^{^{{\scriptscriptstyle(}{\scriptstyle
\sigma}{\scriptscriptstyle)}}} \right) \ \ . \label{II17}
\end{equation}
On the other hand, the condition $\mbox{\boldmath ${E}$}=0 $
implies the equation ${\partial \mbox{\boldmath ${A}$}}/{\partial
t}+c\mbox{\boldmath $\nabla$}A_0=0$ which, after combined with $q
\mbox{\boldmath ${A}$}= c\mbox{\boldmath $\nabla$}{\cal S}$,
becomes $\partial {\cal S}/\partial t + qA_0=0 $. This last
equation, after taking into account the definition of $A_0$ and
Eq. (\ref{II17}), assumes the form
\begin{equation}
\frac{\partial {\cal S}}{\partial t}+\frac{1}{2\mu}\left(
\mbox{\boldmath $\nabla$} {\cal S }-\frac{q}{c}\mbox{\boldmath
${A}$} ^{^{{\scriptscriptstyle(}{\scriptstyle
ex}{\scriptscriptstyle)}}}\!\!\!\!\!-\frac{q}{c}\mbox{\boldmath
${A}$} ^{^{{\scriptscriptstyle(}{\scriptstyle
\sigma}{\scriptscriptstyle)}}} \right)^2\!\!+ qA_0
^{^{{\scriptscriptstyle(}{\scriptstyle
ex}{\scriptscriptstyle)}}}\!\!+ qA_0
^{^{{\scriptscriptstyle(}{\scriptstyle
\sigma}{\scriptscriptstyle)}}}\! =0. \label{II18}
\end{equation}

The vorticity \cite{TA} of a particle system of total mass $m$ and
charge $e$, immersed in an external electromagnetic field, is
defined through $\mbox{\boldmath ${\omega}$}=\mbox{\boldmath
$\nabla$} \times\left(m\mbox{\boldmath ${u}$}+(e/c)\mbox{\boldmath
${A}$} ^{^{{\scriptscriptstyle(}{\scriptstyle
ex}{\scriptscriptstyle)}}}\right)$ and results from the stress
force $\mbox{\boldmath
${F}$}^{^{{\scriptscriptstyle(}{\scriptstyle
\sigma}{\scriptscriptstyle)}}}$, being $\mbox{\boldmath
${\omega}$}=- (e/c) \mbox{\boldmath $\nabla$} \times
\mbox{\boldmath ${A}$}^{^{{\scriptscriptstyle(}{\scriptstyle
\sigma}{\scriptscriptstyle)}}}=-(e/c)\mbox{\boldmath ${B}$}
^{^{{\scriptscriptstyle(}{\scriptstyle
\sigma}{\scriptscriptstyle)}}}$, as one can verify immediately by
considering Eq. (\ref{II17}).

According to ref. \cite{TA} we can decompose the stress tensor
density as $\sigma_{jk}=\! \varsigma_{jk} +\!\nu_{jk} $. The first
term is the residual or nonvortical stress tensor density, which
is present also when $\mbox{\boldmath ${\omega}$}=0$. The second
term represents the vortical stress tensor density, which
originates the vortical flow in the system. From Eq. (\ref{II10})
we have that ${\mbox{\boldmath ${F}$}}
^{^{{\scriptscriptstyle(}{\scriptstyle
\sigma}{\scriptscriptstyle)}}}={\mbox{\boldmath ${F}$}}
^{^{{\scriptscriptstyle(}{\scriptstyle
\varsigma}{\scriptscriptstyle)}}}+{\mbox{\boldmath ${F}$}}
^{^{{\scriptscriptstyle(}{\scriptstyle
\nu}{\scriptscriptstyle)}}}$. Of course it results  $A_0
^{^{{\scriptscriptstyle(}{\scriptstyle
\sigma}{\scriptscriptstyle)}}}\!\!=\!
A_0^{^{{\scriptscriptstyle(}{\scriptstyle
\varsigma}{\scriptscriptstyle)}}}+\!A_0
^{^{{\scriptscriptstyle(}{\scriptstyle
\nu}{\scriptscriptstyle)}}}$, while we can pose $\mbox{\boldmath
${A}$}^{^{{\scriptscriptstyle(}{\scriptstyle
\sigma}{\scriptscriptstyle)}}}=\!\mbox{\boldmath ${A}$}
^{^{{\scriptscriptstyle(}{\scriptstyle
\nu}{\scriptscriptstyle)}}}$ in order to write $\mbox{\boldmath
${\omega}$}=- (e/c) \mbox{\boldmath $\nabla$} \times
\mbox{\boldmath ${A}$}^{^{{\scriptscriptstyle(}{\scriptstyle
\nu}{\scriptscriptstyle)}}}$ and then $\mbox{\boldmath
${\omega}$}=- (e/c) \mbox{\boldmath
${B}$}^{^{{\scriptscriptstyle(}{\scriptstyle
\nu}{\scriptscriptstyle)}}}$. Consequently for the nonvortical
stress force we have
\begin{equation}
{F}_j ^{^{{\scriptscriptstyle(}{\scriptstyle
\varsigma}{\scriptscriptstyle)}}}=-\frac{1}{\rho}\,\frac{\partial
}{\partial x_k}\,\,\rho \, \varsigma_{jk} \ \ ; \ \
\mbox{\boldmath ${F}$} ^{^{{\scriptscriptstyle(}{\scriptstyle
\varsigma}{\scriptscriptstyle)}}}=-q\mbox{\boldmath $\nabla$}
A_0^{^{{\scriptscriptstyle(}{\scriptstyle
\varsigma}{\scriptscriptstyle)}}}    \ \ . \label{II19}
\end{equation}
For the vortical stress force we obtain
\begin{equation}
{F}_j ^{^{{\scriptscriptstyle(}{\scriptstyle
\nu}{\scriptscriptstyle)}}}\!=-\frac{1}{\rho}\,\frac{\partial
}{\partial x_k} \,\rho\,\nu_{jk} \ \ ; \ \ \mbox{\boldmath ${F}$}
^{^{{\scriptscriptstyle(}{\scriptstyle
\nu}{\scriptscriptstyle)}}}\!\!=q \left(\mbox{\boldmath ${E}$}
^{^{{\scriptscriptstyle(}{\scriptstyle
\nu}{\scriptscriptstyle)}}}\!\!+\!\frac{1}{c}\,\mbox{\boldmath
${u}$} \!\times\!\mbox{\boldmath ${B}$}
^{^{{\scriptscriptstyle(}{\scriptstyle
\nu}{\scriptscriptstyle)}}}\right), \label{II20}
\end{equation}
where the fields $\mbox{\boldmath ${E}$}
^{^{{\scriptscriptstyle(}{\scriptstyle
\nu}{\scriptscriptstyle)}}}$ and $\mbox{\boldmath ${B}$}
^{^{{\scriptscriptstyle(}{\scriptstyle
\nu}{\scriptscriptstyle)}}}$ are derived from the potential $A
^{^{{\scriptscriptstyle(} {\scriptstyle
\nu}{\scriptscriptstyle)}}}=(A_0
^{^{{\scriptscriptstyle(}{\scriptstyle \nu}{\scriptscriptstyle)}}}
,\mbox{\boldmath ${A}$} ^{^{{\scriptscriptstyle(}{\scriptstyle
\nu}{\scriptscriptstyle)}}})$ through
\begin{equation}
\mbox{\boldmath ${E}$} ^{^{{\scriptscriptstyle(}{\scriptstyle
\nu}{\scriptscriptstyle)}}}\!\!=-\mbox{\boldmath $\nabla$} A_0
^{^{{\scriptscriptstyle(}{\scriptstyle
\nu}{\scriptscriptstyle)}}}\!-\frac{1}{c}\frac{\partial
\mbox{\boldmath ${A}$} ^{^{{\scriptscriptstyle(}{\scriptstyle
\nu}{\scriptscriptstyle)}}}}{\partial t} \ \ ; \ \ \mbox{\boldmath
${B}$} ^{^{{\scriptscriptstyle(}{\scriptstyle
\nu}{\scriptscriptstyle)}}}\!\! = \mbox{\boldmath $\nabla$} \times
\mbox{\boldmath ${A}$} ^{^{{\scriptscriptstyle(}{\scriptstyle
\nu}{\scriptscriptstyle)}}} \ \ . \label{II21}
\end{equation}

Finally Eq.(\ref{II18}) can be written in a form where the
contributions of the nonvortical and vortical stress potentials
appear separately
\begin{eqnarray}
\frac{\partial {\cal S}}{\partial t}
+\frac{1}{2\mu}\left(\mbox{\boldmath $\nabla$}{\cal S
}-\frac{q}{c}\mbox{\boldmath ${A}$}
^{^{{\scriptscriptstyle(}{\scriptstyle
ex}{\scriptscriptstyle)}}}\!\!\!-\frac{q}{c}\mbox{\boldmath ${A}$}
^{^{{\scriptscriptstyle(}{\scriptstyle
\nu}{\scriptscriptstyle)}}} \right)^2&& \nonumber \\
+ \, \, qA_0 ^{^{{\scriptscriptstyle(}{\scriptstyle \varsigma
}{\scriptscriptstyle)}}} +\, qA_0
^{^{{\scriptscriptstyle(}{\scriptstyle ex
}{\scriptscriptstyle)}}} +
\,qA_0^{^{{\scriptscriptstyle(}{\scriptstyle \nu
}{\scriptscriptstyle)}}} =0&& \ . \label{II22}
\end{eqnarray}
We remark that Eq.s (\ref{II22}) and (\ref{II17}) are completely
equivalent to Eq. (\ref{II8}).

The continuity equation, after taking into account (\ref{II17}),
assumes the following form
\begin{equation}
\frac{\partial \rho}{\partial t} + \mbox{\boldmath $\nabla$}
\cdot\left[ \frac{1}{\mu}\left( \mbox{\boldmath $\nabla$}{\cal S}
-\frac{q}{c}\mbox{\boldmath ${A}$}
^{^{{\scriptscriptstyle(}{\scriptstyle
ex}{\scriptscriptstyle)}}}\!\!-\frac{q}{c}\mbox{\boldmath ${A}$}
^{^{{\scriptscriptstyle(}{\scriptstyle \nu}{\scriptscriptstyle)}}}
\right)\rho\right]=0 \ \ . \label{II23}
\end{equation}

It is important to emphasize that Eq.s (\ref{II22}), (\ref{II23})
and (\ref{II17}) describe the most general hydrodynamics in an
alternative fashion with respect to the usual one, which we find
in the texts of fluid physics. Clearly there are still
considerable degrees of freedom in this hydrodynamics and it
necessitates some constitutive equations. It is perhaps
remarkable that one can obtain these additional equations, as we
will see in the following, within the theory and without making
any extra assumption.

\section{Quantum Evolution Equation}

Eq. (\ref{II19}) has a very transparent physical meaning simply
imposing that the nonvortical stress force $\mbox{\boldmath ${F}$}
^{^{{\scriptscriptstyle(}{\scriptstyle
\varsigma}{\scriptscriptstyle)}}}$ is conservative.  After
introducing the field ${\xi} = \ln \rho$ we can write this
equation under the form
\begin{equation}
q\frac{\partial A_0 ^{^{{\scriptscriptstyle(}{\scriptstyle
\varsigma}{\scriptscriptstyle)}}}}{\partial x_j}- \frac{\partial
\varsigma_{jk}}{\partial x_k}-\varsigma_{jk}\frac{\partial
{\xi}}{\partial x_k}=0 \ \ . \label{III1}
\end{equation}

Note that Eq. (\ref{III1}) is a condition constraining the forms
of $A_0^{^{{\scriptscriptstyle(}{\scriptstyle
\varsigma}{\scriptscriptstyle)}}}$ and $\varsigma_{jk}$ which can
be viewed as two functionals of the field $\xi$. Even though it
can appear that Eq. (\ref{III1}) contains considerable degrees of
freedom, in the following we will show that the particular
structure of this equation restricts strongly the number of its
solutions.

We observe that just as the first term in Eq. (\ref{III1}) also
the second and third term must take the form
$\partial(...)/\partial x_j$. This requirement can be satisfied
by posing $\varsigma_{jk}=\varsigma \, \delta_{jk}$. Note that
this choice is compatible with the symmetry property
$\varsigma_{jk}=\varsigma_{kj}$ imposed by the monad kinetics and
permits to write Eq. (\ref{III1}) as
\begin{equation}
q\frac{\partial A_0 ^{^{{\scriptscriptstyle(}{\scriptstyle
\varsigma}{\scriptscriptstyle)}}}}{\partial x_j}- \frac{\partial
\,\varsigma}{\partial x_j}-\varsigma \frac{\partial
{\xi}}{\partial x_j}=0 \ \ . \label{III2}
\end{equation}
The structure of the third term in Eq. (\ref{III2}) imposes that
$\varsigma$ must be an arbitrary  function of $\xi$ so that we
can write
\begin{equation}
\varsigma \frac{\partial {\xi}}{\partial x_j}=\frac{\partial
}{\partial x_j}\int \! \varsigma \, d \xi \ \ . \label{III3}
\end{equation}
In this way Eq. (\ref{III1}) assumes the following simple  form
\begin{equation}
\frac{\partial }{\partial x_j}\left(qA_0
^{^{{\scriptscriptstyle(}{\scriptstyle
\varsigma}{\scriptscriptstyle)}}}- \varsigma-\int \! \varsigma \,
d \xi \right)=0 \ \ . \label{III4}
\end{equation}
Eq. (\ref{III4}) allows to express $A_0
^{^{{\scriptscriptstyle(}{\scriptstyle
\varsigma}{\scriptscriptstyle)}}}$ in terms of $\varsigma $, which
remains an arbitrary function of the variable $\rho$. Then we can
write the first solution (or class of solutions being
$\varsigma(\rho)$ arbitrary) of Eq. (\ref{III1}) as
\begin{eqnarray}
&&\varsigma_{\,jk}=\varsigma (\rho)\, \delta_{jk} \ \ , \\
\label{III5} &&qA_0 ^{^{{\scriptscriptstyle(}{\scriptstyle
\varsigma}{\scriptscriptstyle)}}}= \varsigma (\rho) + \int \! d
\rho \,\frac{\varsigma (\rho)}{\rho} \,  \ \ . \label{III6}
\end{eqnarray}
This solution  describes the classical Eulerian fluid with density
of internal energy $\varepsilon= \varsigma_{jj}/2$ given by
$\varepsilon=3\,\varsigma (\rho)/2$ and pressure $\pi=\rho \,
\varsigma(\rho)$.

We show now that besides the above classical solution, Eq.
(\ref{III1}) admits a second more interesting and less evident
solution. Again, we recall that all the terms of Eq. (\ref{III1})
must take the form $\partial(...)/\partial x_j$. The particularly
simple structure of the second term suggests to chose
$\varsigma_{jk}=\partial a_k/\partial x_j$. Using the symmetry
property $\varsigma_{jk}=\varsigma_{kj}$ we have that
$a_k=\partial \alpha/\partial x_k$ and then the density of
residual stress tensor assumes the form
\begin{equation}
\varsigma_{jk}=\frac{\partial^2 \alpha}{\partial x_j\partial x_k
}\ \ , \label{III7}
\end{equation}
with $\alpha$ an unknown scalar functional depending on the field
$\xi$. Eq. (\ref{III1}) can be written now as
\begin{equation}
\frac{\partial}{\partial x_j}\left(
qA_0^{^{{\scriptscriptstyle(}{\scriptstyle
\varsigma}{\scriptscriptstyle)}}} -\frac{\partial^2 \alpha
}{\partial x_k\partial x_k}\right) - \frac{\partial^2 \alpha
}{\partial x_j\partial x_k} \, \frac{\partial {\xi}}{\partial
x_k} =0 \ \ . \label{III8}
\end{equation}

By making the hypothesis that the functional $\alpha$ is a
function of the field ${\xi}$ (it can be easily verified that
this is the only possibility), and after developing the
derivatives of $\alpha(\xi)$ we obtain
\begin{eqnarray}
\frac{\partial^2 \alpha }{\partial x_j\partial x_k} = \frac{d
{\alpha}}{d{\xi}} \frac{\partial^2{\xi}}{\partial x_j\partial x_k}
\!+ \!\! \frac{d^2 {\alpha}}{d{\xi}^2} \frac{\partial
{\xi}}{\partial x_j} \frac{\partial {\xi}}{\partial x_k} \ \ .
\label{III9}
\end{eqnarray}
Consequently, the third term in Eq.(\ref{III8}) becomes
\begin{eqnarray}
\frac{\partial^2 \alpha }{\partial x_j\partial x_k} \,
\frac{\partial {\xi}}{\partial x_k} \,\,&&
=\frac{\partial}{\partial x_j}\left[ \frac{1}{2} \frac{d
{\alpha}}{d{\xi}} \frac{\partial{\xi}}{\partial x_k}
\frac{\partial {\xi}}{\partial x_k} \right] \nonumber \\ && +\,
\frac{1}{2}\, \frac{d^2 {\alpha}}{d{\xi}^2}\,
 \frac{\partial {\xi}}{\partial x_j}\,
\frac{\partial {\xi}}{\partial x_k} \frac{\partial
{\xi}}{\partial x_k} \ \ . \label{III10}
\end{eqnarray}
The requirement that also this third term has the form
$\partial(...)/\partial x_j$, imposes that $d^2 \alpha /d
{\xi}^2=0$ so that Eq. (\ref{III1}) assumes the form
\begin{equation}
\frac{\partial}{\partial x_j}\left[
qA_0^{^{{\scriptscriptstyle(}{\scriptstyle
\varsigma}{\scriptscriptstyle)}}} -  \frac{d {\alpha}}{d{\xi}}
\frac{\partial^2{\xi}}{\partial x_k\partial x_k} - \frac{1}{2}
\frac{d {\alpha}}{d{\xi}}\frac{\partial{\xi}}{\partial
x_k}\frac{\partial {\xi}}{\partial x_k}\right] =0 \ \ .
\label{III11}
\end{equation}

The condition $d^2 \alpha /d {\xi}^2=0$ is a second order ordinary
differential equation which can be trivially integrated providing
$\alpha=c_0\xi+ c_1$ with $c_0$ and $c_1$ arbitrary integration
constants. The expressions of $\varsigma_{jk}$ and
$A_0^{^{{\scriptscriptstyle(}{\scriptstyle
\varsigma}{\scriptscriptstyle)}}}$, follow immediately from
Eq.(\ref{III7}) and Eq. (\ref{III11}) respectively
\begin{eqnarray}
&& \varsigma_{jk}=c_0\,\,\frac{\partial^{\,2} }{\partial x_j
\,\partial x_k} {\,\ln \rho} \label{III12} \ \ , \\ && qA_0
^{^{{\scriptscriptstyle(}{\scriptstyle
\varsigma}{\scriptscriptstyle)}}}= c_0 \left[ \, \Delta \ln \rho
+\frac{1}{2} \left( \mbox{\boldmath $\nabla$} \ln \rho
\right)^{2} \right ] \ \ . \label{III13}
\end{eqnarray}
The constant $c_1$, not influencing the values of
$\varsigma_{jk}$ and $A_0^{^{{\scriptscriptstyle(}{\scriptstyle
\varsigma}{\scriptscriptstyle)}}}$, can be set equal to zero.

From the definition of the total internal energy related to
nonvortical flow we have ${\cal H}=\int d^3x \, \rho \, \,
\varepsilon > 0$ being $\varepsilon =\varsigma_{jj}/2
>0$ its density \cite{GL,GK}. Starting from Eq. (\ref{III12})
we can easily calculate  ${\cal H}$ obtaining
\begin{equation}
{\cal H}=\frac{\eta^2}{8 \mu}\,\,I \ \ \ ; \ \ \ I=\int d^3x
\,\frac{1}{\rho}\,(\mbox{\boldmath $\nabla$} \rho)^2 \ \ ,
\label{III14}
\end{equation}
where we have posed $c_0=-\eta^2/4\mu<0$ in order to have ${\cal
H}> 0$. The real positive constant $\eta$ remains a free
parameter of the theory. It is remarkable that ${\cal H}$ turns
out to be proportional to the Fisher information measure $I$
(Fisher 1922) \cite{RE,FR} which in this way, emerges in quantum
mechanics naturally.

Note that after setting $\eta=\hbar/N$ the quantity
$eA_0^{^{{\scriptscriptstyle(}{\scriptstyle
\varsigma}{\scriptscriptstyle)}}}$ given by Eq. (\ref{III13})
results to be the quantum potential (Madelung 1926) \cite{MA,BO}.
In this way, the fundamental constant $\hbar$ comes out simply as
an integration constant, while the quantum potential emerges
naturally as one of the two possible solutions of Eq.
(\ref{III1}).

The two solutions above obtained are the only possible ones. At
the present, we do not have any criterion to judge whether one
solution or the other is the right solution to be chosen as
constitutive equation for the system. It is trivial to verify
that the most general solution of Eq. (\ref{III1}) can be
expressed as a linear combination of these independent solutions
and assumes the form
\begin{eqnarray}
&&\!\!\!\!\!\!\!\!\!\! \varsigma_{jk}=\delta_{jk}
\frac{1}{\rho}\!\int\!\rho \frac{d U(\rho)}{d\rho}d\rho \,-
\frac{\hbar^2}{4m} \,\,\frac{\partial^{\,2} }{\partial x_j
\,\partial x_k} {\,\ln \rho} \ \ , \label{III15}
\\ && \!\!\!\!\!\!\!\!\!\! eA_0^{^{{\scriptscriptstyle(}{\scriptstyle
\varsigma}{\scriptscriptstyle)}}}=U(\rho)
-\frac{\hbar^2}{4m}\left[ \Delta \ln \rho +\frac{1}{2} \left(
\mbox{\boldmath $\nabla$} \ln \rho \right)^{2} \right ] \ \ ,
\label{III16}
\end{eqnarray}
being $U(\rho)$ an arbitrary function. Eq. (\ref{III16})
represents the wanted constitutive equation and has been enforced
exclusively from the fact that the nonvortical component of
stress force is conservative.

Note that the constitutive equation (\ref{III16}), which gives the
potential $eA_0^{^{{\scriptscriptstyle(}{\scriptstyle
\varsigma}{\scriptscriptstyle)}}}$ starting from $\rho$, has been
obtained in the framework of an entirely classical kinetics. This
kinetics describes the subquantum statistical ensemble  of the
$N$ interacting monads. The monad interaction, which is not
specified, generates the stress forces and then the potential
$eA_0^{^{{\scriptscriptstyle(}{\scriptstyle
\varsigma}{\scriptscriptstyle)}}}$. Clearly only when the monad
interactions are suppressed  the potential vanishes. Then the
presence of the collision integral $C(f)$ in the Eq. (\ref{II2})
is necessary for the consistency of the theory. Concerning the
potential $eA_0^{^{{\scriptscriptstyle(}{\scriptstyle
\varsigma}{\scriptscriptstyle)}}}$ we note that in the limit
$\hbar \rightarrow 0$, becomes
$eA_0^{^{{\scriptscriptstyle(}{\scriptstyle
\varsigma}{\scriptscriptstyle)}}}=U(\rho)$. In this case the
constitutive equation describes a classical fluid which is
governed by Eq.s (\ref{II22}) and (\ref{II23}).

Observe that the system can be described through the two real
fields $\rho$ and ${\cal S}$, whose evolution equations are
(\ref{II23}) and (\ref{II22}), respectively. Obviously, these
equations must be considered together with the constitutive
equation (\ref{III16}) that defines
$A_0^{^{{\scriptscriptstyle(}{\scriptstyle
\varsigma}{\scriptscriptstyle)}}}$. Alternatively one can describe
the system by means of the complex field
\mbox{$\Psi=\rho^{1/2}\exp \left(i S/\hbar\right)$} with $S=N{\cal
S}$. The evolution equation for the field $\Psi$ can be obtained
trivially by using the standard procedure \cite{GK}, starting from
Eq.s (\ref{II22}), (\ref{II23}) and (\ref{III16})
\begin{eqnarray}
i\hbar\frac{\partial \Psi}{\partial t} =&& \!\!\!\!
\frac{1}{2m}\left(-i{\hbar}\mbox{\boldmath
$\nabla$}-\frac{e}{c}\mbox{\boldmath ${A}$}
^{^{{\scriptscriptstyle(}{\scriptstyle
ex}{\scriptscriptstyle)}}}\!\!\!-\frac{e }{c}\mbox{\boldmath
${A}$} ^{^{{\scriptscriptstyle(}{\scriptstyle
\nu}{\scriptscriptstyle)}}} \right)^2\!\Psi \nonumber
\\ \! \!&& +\,\,  \,U(\rho)\,\Psi +\, eA_0
^{^{{\scriptscriptstyle(}{\scriptstyle
ex}{\scriptscriptstyle)}}}\Psi + eA_0
^{^{{\scriptscriptstyle(}{\scriptstyle \nu}{\scriptscriptstyle)}}}
\Psi \ . \label{III17}
\end{eqnarray}
Eq. (\ref{III17}) can describe the non relativistic quantum
particle with spin in an external electromagnetic field generated
from the potential $A ^{^{{\scriptscriptstyle(} {\scriptstyle
ex}{\scriptscriptstyle)}}}=(A_0
^{^{{\scriptscriptstyle(}{\scriptstyle ex}{\scriptscriptstyle)}}}
,\mbox{\boldmath ${A}$} ^{^{{\scriptscriptstyle(}{\scriptstyle
ex}{\scriptscriptstyle)}}})$.  The second potential $A
^{^{{\scriptscriptstyle(} {\scriptstyle
\nu}{\scriptscriptstyle)}}}=(A_0
^{^{{\scriptscriptstyle(}{\scriptstyle \nu}{\scriptscriptstyle)}}}
,\mbox{\boldmath ${A}$} ^{^{{\scriptscriptstyle(}{\scriptstyle
\nu}{\scriptscriptstyle)}}})$ represents internal degrees of
freedom and describes the vortical flow of the system with
$\mbox{\boldmath ${\omega}$}\neq 0$. Finally in Eq. (\ref{III17})
a third, internal, arbitrary and nonlinear potential namely
$U(\rho)$ appears, which can be exploited to investigate some
complex phenomenologies of the condensed matter arising from
collective interactions (e.g. the Bose-Einstein condensation has
been studied previously in the literature by considering
$U(\rho)=a\rho$).

Besides the equation of motion the present theory is
characterized by the presence of a subsidiary condition
restricting the velocity circulation. To show this we rewrite
Eq.(\ref{II17}) under the form
\begin{equation}
m\,\mbox{\boldmath ${u}$}+\frac{e}{c}\mbox{\boldmath ${A}$}
^{^{{\scriptscriptstyle(}{\scriptstyle
ex}{\scriptscriptstyle)}}}\!\!\!+\frac{e}{c}\mbox{\boldmath ${A}$}
^{^{{\scriptscriptstyle(}{\scriptstyle
\sigma}{\scriptscriptstyle)}}} =  \mbox{\boldmath $\nabla$} {S} \
\ . \label{III18}
\end{equation}
Consider an arbitrary closed contour $C$ delimiting the surface
$s$, it is trivial to verify that from Eq. (\ref{III18}) follows
\begin{equation}
m\oint_C\mbox{\boldmath ${u}$}\,d\mbox{\boldmath
${l}$}+\frac{e}{c}\Phi-\frac{\hbar}{2}\int_s T_{_N} ds = \Gamma \
\ , \label{III19}
\end{equation}
being $\Phi=\int_s B^{^{{\scriptscriptstyle(}{\scriptstyle
ex}{\scriptscriptstyle)}}}_{_N} ds$ the flux of the external
magnetic field $\mbox{\boldmath ${B}$}
^{^{{\scriptscriptstyle(}{\scriptstyle ex}{\scriptscriptstyle)}}}$
going through $s$, while the vector $\mbox{\boldmath ${T}$}$ is
proportional to the vorticity namely $\mbox{\boldmath
${\omega}$}=(\hbar/2)\mbox{\boldmath ${T}$}$. Then if we take
into account that $\mbox{\boldmath ${\omega}$}=- (e/c)
\mbox{\boldmath ${B}$}^{^{{\scriptscriptstyle(}{\scriptstyle
\nu}{\scriptscriptstyle)}}}$ we have that the third term in the
left hand side of Eq. (\ref{III19}) is proportional to the flux
of the $\mbox{\boldmath ${B}$}
^{^{{\scriptscriptstyle(}{\scriptstyle
\nu}{\scriptscriptstyle)}}}$ going through $s$. Finally the
quantity $\Gamma$ is given by
\begin{equation}
\Gamma =\oint_C \mbox{\boldmath $\nabla$} {S } \,d\mbox{\boldmath
${l}$}= \oint_C d{S } \ \ , \label{III20}
\end{equation}
and its value does not depend on the detailed path of $C$ in so
far as it does not pass through a singular line because
$\mbox{\boldmath$\nabla$}\times\mbox{\boldmath$\nabla$}{ S}=0$
elsewhere.  We remark that the condition (\ref{III19}) holds
independently on the spin value which clearly influences the
value of $\Gamma$. For spinless particles $(\mbox{\boldmath
${T}$}=0)$ it has been shown that $\Gamma=nh$ with $n$ an
arbitrary integer \cite{GK,TA}. The case of a non relativistic
particle with spin $1/2$ has been considered extensively in the
framework of the Schr\"odinger-Pauli theory, by several authors
\cite{TA,SO,SI,GH}, and it has been shown that $\Gamma=nh/2$
\cite{TA}.

\section{Concluding Remarks}

We tackle briefly the problem concerning the locality in quantum
physics, that was left unresolved in the twenty three year long
debate between Einstein and Bohr and was reconsidered by Bell in
1964. It is well known that the Bell's inequality has been
obtained in the framework of local, hidden variables and
deterministic theories. This inequality is in disagreement both
with quantum mechanics and experimental evidence. The reason of
this disagreement now appears clear. Here we have obtained quantum
mechanics starting from the underlying monad kinetics which is a
non local, hidden variables and probabilistic theory. Indeed, in
the proposed scenario, the quantum particle of mass $m$ turns out
to be a statistical system, having a spatial extension and an
internal structure. It is composed by $N$ identical subquantum
interacting particles of mass $\mu$, the monads. These monads
obey the laws of classical physics and their dynamics is
described in the phase space by the standard kinetic equation.

Remark that we don't make any assumption about the structure of
the collision integral and the nature of the interaction between
the monads except that, during the point-like collisions the monad
number, momentum and energy are conserved. The projection of the
above phase space kinetics into the physical space produces a
hydrodynamics which leads naturally to the one particle quantum
mechanics. On the other hand the monad interaction generates the
stress forces and then the quantum potential which vanishes only
when the monad interactions are suppressed. Then the presence of
the collision integral in the kinetic Eq. (\ref{II1}) is
necessary for the consistency of the theory. The determination of
the form of the collision integral in the framework of a
subquantum dynamics is a very interesting problem which remains
still open.

At this point, the question arises spontaneously if it is
possible to include the locality in quantum physics. Clearly,
there is no problem in principle to construct a theory starting
from a relativistic kinetic equation rather than from Eq.
(\ref{II2}). After noting that a subquantum relativistic kinetics
is a local, hidden variables and probabilistic theory, we can
make the conjecture that {\it a subquantum relativistic monad
kinetics could be underlying a local quantum theory}.

To summarize we have shown that the approach to quantum mechanics
proposed in ref. \cite{GK} is suitable to treat also the quantum
particle with spin as well as the nonlinear quantum mechanics.
Main goal of the present paper is the quantum evolution equation
(\ref{III17}) which has been obtained within an entirely
classical subquantum kinetics without making any extra
assumption. This equation: i) can describe the quantum particle
with spin, due to the presence of the internal potential $A
^{^{{\scriptscriptstyle(} {\scriptstyle
\nu}{\scriptscriptstyle)}}}$; ii) can be used to study some
complex phenomenologies in condensed matter physics e.g.
Bose-Einstein condensation, due to the presence of the
nonlinearity $U(\rho)$ . Clearly when $A
^{^{{\scriptscriptstyle(} {\scriptstyle
\nu}{\scriptscriptstyle)}}}=0$ and $U(\rho)=0$, Eq. (\ref{III17})
reproduces the Schr\"odinger equation describing the spinless
particle.

\end{document}